\documentstyle[prd,aps,floats,epsfig]{revtex}
\newcommand{\be}{\begin{eqnarray}}
\newcommand{\ee}{\end{eqnarray}}
\newcommand{\real}{\mbox{{\rm I\hspace{-2truemm} R}}}
\newcommand{\im}{\mbox{{\rm I\hspace{-1.6truemm} I}}}
\newcommand{\realp}{\mbox{\real{\rm e}\,}}
\newcommand{\imp}{\mbox{\im{\rm m}\,}}
\newcommand{\bol}[1]{\mbox{\boldmath $#1$}}
\newcommand{\id}{\mbox{{\rm 1\hspace{-2truemm} I}}}
\newcommand{\ps}{\mbox{$\not\hspace{-0.5truemm}\partial$}}
\newcommand{\Ds}{\mbox{$\not\hspace{-1truemm}D$}}
\newcommand{\As}{\mbox{$\not\hspace{-1truemm}A$}}

\newcommand{\ad}{\mbox{$\hat a^\dagger$}}
\renewcommand{\a}{\mbox{$\hat a$}}
\begin{document}
\draft
\twocolumn[\hsize\textwidth\columnwidth\hsize\csname
@twocolumnfalse\endcsname
\title{Discrete Symmetries and Localization in a Brane-world}
\author{R. Casadio$^{a}$ and A. Gruppuso$^{b}$}
\address{~}
\address{Dipartimento di Fisica, Universit\`a di Bologna,
and I.N.F.N., Sezione di Bologna \\
via Irnerio 46, I-40126 Bologna, Italy}
\date{\today}
\maketitle
\begin{abstract}
Discrete symmetries are studied in warped space-times with one extra
dimension.
In particular, we analyze the compatibility of five- and four-dimensional
charge conjugation, parity, time reversal and the orbifold symmetry
Z$_2$ with localization of fermions on the four-dimensional brane-world
and Lorentz invariance.
We then show that, when a suitable topological scalar field (the ``kink'')
is included, fermion localization is a consequence of (five-dimensional)
CPT invariance.
\end{abstract}
]
\section{Introduction}
\setcounter{equation}{0}
Recently there has been a revived interest in models containing
extra spatial dimensions.
The first time this idea was put in concrete form possibly dates
back to the '20s \cite{kaluza}, when the existence of more than
four dimensions was employed in an attempt to unify gravity with
the electromagnetic field.
The lack of signals from (and loss into) the fifth dimension was
assured by taking the size of the extra dimension small, which
gives a huge mass gap between the ground state fields and the
Kaluza-Klein (KK) modes.
One could then Fourier-transform all fields, retain just the
lowest mode and integrate out any dependence on the fifth
dimension.
In the context of such theories, the discrete symmetries of
charge conjugation (C), parity (P) and time reversal (T) were
investigated in Ref.~\cite{thirring}, where it was found that
the Dirac equation contains terms which violate CP and T.
However, such violating terms are highly suppressed by factors
of $1/M_p$, where $M_p$ is the Planck mass, and are not likely
to lead to any observable effect.
\par
Higher dimensional spaces naturally arise in string theory
\cite{green}, where one must compactify the fundamental theory
down to our four-dimensional world.
In Ref.~\cite{horava} a compactification mechanism was built
for the strongly coupled $E_8\times E_8$ heterotic string,
whose low energy limit is eleven-dimensional supergravity.
Six dimensions are then compactified on a Calabi-Yau
manifold and integrated out to leave a five-dimensional
(bulk) space-time bounded by two copies of the same D3-brane
\cite{polchinski}.
Matter particles are low energy excitations of open strings,
whose end-points are tight to the D3-brane, which could thus
represent our (brane-) world.
Gravitons are instead closed strings and can propagate also in
the extra direction, which has topology $S^1/$Z$_2$.
Further, this construction yields a relation between the Planck
scale on the D3-brane and the fundamental string scale which
allows the latter to be much smaller than the former.
\par
Starting from this result, several models have been proposed,
with a variety in the number of extra dimensions, which can
further be either compact \cite{arkani} or infinitely
extended but with a warp factor along the extra directions
\cite{randall}.
In both cases, the parameters can be tuned in such a way that
the fundamental mass scale $M$ is small enough to lead to new
physics slightly above 1 TeV without violating Newton's law
at the present level of confidence \cite{long}.
One of the main concerns in such models is actually to provide an
explicit, low energy, confining mechanism for the matter fields
which does not violate any of the tested properties of the
Standard Model and yields, at the same time, predictable
effects to be probed by the forthcoming generation of detectors
\cite{arkani,giudice}.
Early proposals for confining matter fields on a four-dimensional
wall are actually older (see Refs.~\cite{rubakov}) and make use
of a mechanism which was proposed to generate masses dynamically
by coupling fermions to a solitonic state (``kink'') of a scalar
field.
Such a mechanism yields a brane whose thickness
\cite{schmaltz,lykken} can then be related to the existence of
KK-like partners of the SM particles and constrained by comparing
with precision measurements such as the anomalous magnetic moment
of leptons \cite{g2}.
The fact that bulk gravitons living in the extra dimensions
have not yet been detected is then generally a consequence of the
small coupling with SM particles, namely $1/M_p$
\cite{arkani,graesser}.
\par
In the present paper we analyze general aspects of the discrete
symmetries C, P and T for fermions interacting with gauge fields and
the Z$_2$ symmetry along the extra dimension in much the same spirit
as such symmetries were studied in Ref.~\cite{thirring}.
(For the sake of simplicity, we just consider one extra dimension
and abelian gauge fields.)
In particular, we are interested in the interplay between CPT (and
Lorentz invariance) in five and four dimensions and the mechanism
of confinement for fermions
mentioned above from the viewpoint of local field theory, without
attempting any rigorous connection to string theory or D-brane
theory.
\par
In the next Section we define the model.
In Section~\ref{discrete} we introduce C, P and T both from the
five- and four-dimensional points of view and highlight some
ambiguities in the definition of the corresponding
transformations.
However, we show that there is a unique form of five-dimensional CPT
invariance which is compatible with localization.
Finally, in Section~\ref{conclusions} we summarize and comment
on our results.
\section{Localized fermions and fermion doubling}
We consider a five-dimensional space-time with (non-factorizable)
metric $g_{AB}$ ($A,B=0,...,4 \,$) given by
\be
ds^2=a^2\,d{\bol x}\cdot d{\bol x}+dy^2
\ ,
\label{m}
\ee
where $\cdot$ is the flat (Minkowski) scalar product between
four-dimensional vectors.
We do not assume for the warp factor $a=a(y)$ any given form.
In principle, $a$ could be determined once the matter content is
given by solving Einstein field equations, however, in this
article we shall just need to know that it is Z$_2$-even,
to wit $a(-y)=a(y)$.
The four-dimensional metric on the brane, which is located
at $y=0$, is $\bol{\eta}={\rm diag}\,[-1,+1,+1,+1]$ and the four
brane coordinates are denoted by ${\bol x}=(t,\vec x)$.
In order to describe spinors, we introduce the ``pentad''
\be
e_{\ B}^A=
{\rm diag}\,\left[{1\over a},{1\over a},{1\over a},{1\over a},
1\right]
\ .
\ee
All fields will be taken as functions of $({\bol x},y)$ and
Dirac matrices satisfy the five-dimensional Clifford algebra
(note the appearance of the warp factor $a$)
\be
e_{\ C}^A\,e_{\ D}^B\,\left\{\gamma^C,\gamma^D\right\}=-2\,g^{AB}
\ ,
\ee
and are given in the Weyl representation by
\be
\gamma^0=\left[\begin{array}{cc}
0 & \id_2 \\
\id_2 & 0\end{array}\right]
\ ,\ \ \
\vec\gamma=\left[\begin{array}{cc}
0 & -\vec\sigma \\
\vec\sigma & 0\end{array}\right]
\ ,
\ee
with $\vec\sigma$ the Pauli spin matrices, and
$\gamma^4=i\,\gamma^5=i\,{\rm diag}\,[+\id_2,-\id_2]$,
where $\id_2$ is the $2\times 2$ identity matrix.
\par
The five-dimensional action for one Dirac spinor
$\psi=\psi(\bol{x},y)$ minimally coupled to a gauge boson
$A_A=A_A(\bol{x},y)$ in the background with metric (\ref{m})
is given by
\be
S=\int a^4\,dy\,\int d^4x\,\bar\psi\left[
{i\over a}\Ds+i\,\gamma^4\left(D_y+\omega\right)
-V\right]\psi
\ ,
\label{5Daction1}
\ee
where $D_y=\partial_y+i\,e\,A_4$, $\Ds=\ps+i\,e\,\As$,
$\ps={\bol \gamma}\cdot{\bol\partial}$,
$\As={\bol \gamma}\cdot{\bol A}$ and $\omega=\partial_y \ln(a^2)$
is a sum of spin connection terms.
In the following Dirac spinor fields are taken in the Weyl
representation,
\be
\psi=\left(\begin{array}{c}
\psi_L \\
\psi_R
\end{array}
\right)
\ ,
\ee
with $L$ for left- and $R$ for right-handed, and their field
equation reads
\be
\left[{i\over a}\,\Ds
+i\,\gamma^4\,\left(\partial_y+ieA_4+\omega\right)
-V\right]\psi=0
\ .
\label{D1}
\ee
The potential $V=m+\Phi$, where the constant $m$ is the
five-dimensional mass of the spinor and $\Phi=\Phi(\bol{x},y)$ is
a (complex) scalar field.
\par
The (orbifold) symmetry Z$_2:\ y\to -y$ requires that the
Lagrangian density appearing in the action for any
given set of fields be an even function of $y$.
If we then assume $S$ be Z$_2$ invariant, one finds that
$\psi_L$ and $\psi_R$ must have opposite parity under $y\to -y$
and the potential $V$ must be Z$_2$-odd, that is
\be
&m=0&
\nonumber \\
\label{cZ2} \\
&\Phi({\bol x},y)=-\Phi({\bol x},-y)&
\ .
\nonumber
\ee
\par
An example of scalar field $\Phi$ which is used to confine
fermions on a domain wall is that of the kink ($+$) or
anti-kink ($-$) \cite{kink,rubakov},
\be
\Phi=\pm \, q\,\tanh\,(p\,y)
\ .
\label{kink}
\ee
The above (positive) parameters $p$ and $q$ are related
to the thickness of  the brane and energy threshold of confinement,
as discussed in Ref.~\cite{schmaltz}.
There it is shown that, on properly taking into account the finite
thickness of the brane, the field $\psi$ can be expanded in a tower
of KK-like modes.
Here, for simplicity, we consider infinitesimal the thickness of the
brane and set $p\to\infty$ in (\ref{kink}).
This yields a scalar field
\be
\Phi\sim \pm \, q\,\mathrm{sgn}(y)
\ ,
\label{kink2}
\ee
which confines just {\em one\/} of the chiralities of a Dirac spinor.
This fact prevents a mass term for the localized spinor and,
as a consequence, we cannot interpret the confined fermions as
the usual standard model particles.
This problem is also present in the case of a thick brane but just
for the zero-mode of the KK tower (see Appendix~\ref{thick} for
details).
\par
A solution to the above problem is the ``fermion doubling'':
since we need both chiralities to recover the correct low energy
phenomenology, one doubles the fermion fields $\psi\to(\psi_1,\psi_2)$
and couples $\psi_1$ to the kink [plus sign in Eq.~(\ref{kink})] and
$\psi_2$ to the ``anti-kink'' [minus sign in Eq.~(\ref{kink})].
It is useful to organize the two kind of fermions in a single
field \cite{rubakov2}
\be
\Psi\equiv
\left(\begin{array}{c}
\psi_{1} \\
\psi_{2} \\
\end{array}
\right)  \equiv
\left(\begin{array}{c}
\psi_{1,L} \\
\psi_{1,R} \\
\psi_{2,L} \\
\psi_{2,R} \\
\end{array}\right)
\ ,
\label{doubling}
\ee
where $\psi_{1}$ and $\psi_{2}$ are two four-component spinors living
in five dimensions.
Further, because of the Z$_2$ symmetry, $\psi_{1,L}$ and $\psi_{2,R}$
must be even (odd) functions of $y$, while $\psi_{1,R}$ and $\psi_{2,L}$
must be odd (even).
After the fermion doubling, the five-dimensional action
(\ref{5Daction1}) can be written as
\be
S_{(5)}&=&\int a^4\,dy\,\int d^4x\,\bar\Psi\,\left\{
{i\over a}\,\left[\bol{\gamma}\otimes\id_2\right]\cdot\bol{D}\right.
\nonumber \\
&&\left.
+i\,\left[\gamma^{4}\otimes\id_2 \right]\,
\left(D_y+\omega\right)-\Phi\,\left[\id_4\otimes\sigma^3\right]
\right.
\nonumber \\
&&\left.\phantom{\vec a\over b}
-{\mu\over a}\,\left[\id_4\otimes\sigma^1\right]\right\}
\,\Psi
\ ,
\label{5Daction+doubling}
\ee
where we have denoted by $\id_4$ the $4\times 4$ identity matrix.
The last term is an interaction term between the two families ($\psi_1$ and
$\psi_2$) of fermions which can give rise to a four-dimensional mass.
In fact, on factorizing the spinor $\psi_i=\alpha(y)\,f_i(\bol{x})$
and first imposing the equation for the localization,
\be
(D_y+\omega+\Phi)\,\alpha=0
\ ,
\label{localization}
\ee
and then the vanishing condition for half of the components,
\be
f_{2,L}=f_{1,R}=0
\ ,
\label{vanishing}
\ee
one obtains the usual four-dimensional action
(multiplied by a numerical coefficient)
\be
S_{(4)}=\int \left|\alpha\right|^2\,a^3\,dy\,\int d^4x\,
\bar\chi\,\left[i\,\Ds-\mu\right]\,\chi
\ .
\label{reduced}
\ee
for the spinor field
\be
\chi=\left(
\begin{array}{c}
f_{1,L} \\
f_{2,R} \\
\end{array} \right)
\label{chi}
\ee
of mass $\mu$.
Clearly the above reduction is successful if the integration on $y$ is
convergent.
Note that the spinor so obtained ($ \chi $) is organized in such a
way that its $L$ component is the $L$ component of the family number
one and its $R$ component is the $R$ component of the family
number two.
Note also that we have considered the gauge field
$A_A$ as independent of $y$.
\par
From the variation of the action in Eq.~(\ref{5Daction+doubling})
we obtain the following equation of motion
\be
&&\left\{{i\over a}\,\left[\bol{\gamma}\otimes\id_2\right]\cdot\bol{D}
+i\left[\gamma^{4}\otimes\id_2\right]\,
\left(D_y+\omega \right)
-\Phi\,\left[\id_4\otimes\sigma^3\right]
\right.
\nonumber \\
&&\left.\phantom{\vec a\over b}
-{\mu\over a}\,\left[\id_4\otimes\sigma^1\right]\right\}\,
\Psi =0
\label{dirac+doubling}
\ .
\end{eqnarray}
The gauge field $A_A$ \cite{gauge} couples minimally
to the fermions via the ``electric'' charge $e$.
We shall not need to work out the field equations satisfied by
$A_A$, so that in the following it can be
taken as a generic external field.
Of course, from the usual parity properties of the electromagnetic tensor
$F_{AB}=\partial_A A_B-\partial_B A_A$, one obtains that the vector potential
must satisfy the following relations
\be
&&A_0(t,\vec x,y)=A_0(-t,\vec x,y)
\nonumber \\
&&\phantom{A^4(t,\vec x,y)}=A_0(t,-\vec x,y)
\nonumber \\
&&\phantom{A^4(t,\vec x,y)}
=A_0(t,\vec x,-y)
\nonumber \\
\nonumber \\
&&\vec A(t,\vec x,y)=-\vec A(-t,\vec x,y)
\nonumber \\
&&\phantom{A^4(t,\vec x,y)}=-\vec A(t,-\vec x,y)
\nonumber \\
&&\phantom{A^4(t,\vec x,y)}
=\vec A(t,\vec x,-y)
\label{Atrans} \\
\nonumber \\
&&A_4(t,\vec x,y)=-A_4(-t,\vec x,y)
\nonumber \\
&&\phantom{A_4(t,\vec x,y)}= A_4(t,-\vec x,y)
\nonumber \\
&&\phantom{A_4(t,\vec x,y)}
=-A_4(t,\vec x,-y)
\nonumber
\ .
\ee
\section{Discrete symmetries}
\label{discrete}
We are now ready to investigate the discrete symmetries Z$_2$,
C, P and T.
As it will be clear from the following, such symmetries are no more
uniquely represented when a fifth dimension is introduced.
In particular, since we are dealing with an effective
five-dimensional field theory, one could define discrete symmetries
from the purely five-dimensional point of view
(see Section~\ref{five}) and consider that CPT holds in five dimensions,
regardless of fermion doubling and thus for each family $\psi_i$
separately.
However, in so doing, one obtains something which does not relate
to the four-dimensional description in a {\em a priori} clear way,
since the fermion doubling mixes components of the two families, and
new definitions of C, P, and T are needed (Section~\ref{four}).
\par
It is our main aim to clarify the compatibility of the above two
aspects and their relation with localization in general.
For our discussion, the precise shape of $a$ and $\Phi$ in the
bulk is not required and we shall mostly need to know just their
parity properties.
\subsection{Five-dimensional symmetries}
\label{five}
In this Section we define discrete symmetries as acting on
each one of the two families $\psi_i$ separately.
These are expected to be the fundamental symmetries of the
five-dimensional effective field theory.
Having grouped together the two families $\psi_i$ according to
Eq.~(\ref{doubling}), we now have to deal with eight-spinors $\Psi$.
Correspondingly, the operators generating (local) Lorentz
transformations as well as C, P and T will be represented
by $8\times 8$ matrices of the form $F\otimes \id_2$,
where $F$ is $4\times 4$.
We shall just consider candidates of C, P and T which have
previously appeared in the literature.
\subsubsection{Charge conjugation}
The first possible definition of five-dimensional charge conjugation
(C$_5$) which we consider is the usual charge conjugation (C)
\cite{peskin} acting on each family $\psi_i$,
\be
C_5\,\left[\Psi \right](x^A)=\left[C \otimes\id_2\right]\,
\Psi^\star(x^A)
\ ,
\label{C5}
\ee
where $C=i\,\gamma^2$ and satisfies
\be
&C& \left(\gamma^{\mu}\right)^{\star}=-\,\gamma^{\mu}\,C
\nonumber  \\
&&      \\
&C& \left( \gamma ^{4} \right) ^{\star} = \, \gamma ^{4} C
\nonumber  \ .
\end{eqnarray}
Invariance of Eq.~(\ref{dirac+doubling}) requires
\be
\left\{i\left[\gamma^4\otimes\id_2\right]
\left(D_y+\omega\right)
-\left[\id_4\otimes\sigma^3\right]\,\imp(\Phi)
\right\}\Psi=0
\ .
\label{cC5}
\ee
If we choose a scalar field of the type in Eq.~(\ref{kink}), we
do not have compatibility with Eq.~(\ref{localization}) for
localization of spinors because the kink is a real function.
It then follows that localization breaks C$_5$.
However, if $\Phi$ is not real, the above condition (\ref{cC5})
is not invariant under rotations and boosts involving the fifth
dimension, therefore it breaks five-dimensional Lorentz invariance
(but is still compatible with four-dimensional Lorentz symmetry).
\par
We can alternatively define charge conjugation in the following way
\cite{pais}:
\be
\tilde C_5\,\left[\Psi \right](x^A)
=\left[\tilde C \otimes\id_2\right]\,
\Psi^\star(x^A)
\ee
where $\tilde C=i\,\gamma^2\,\gamma^5$ and satisfies
\be
\tilde C \left( \gamma ^{A} \right) ^{\star} = \, \gamma ^{A} \tilde C
\ .
\ee
Invariance of Eq.~(\ref{dirac+doubling}) requires
\be
\left\{\realp{\left( \Phi \right)}\,\left[\id_4\otimes\sigma^3\right]
+{\mu\over a}\,\left[\id_4\otimes\sigma^1\right]\right\}\,
\Psi  = 0
\label{condzero}
\ .
\ee
This is satisfied for every $\Psi$ if
\be
\left\{
\begin{array}{l}
\mu = 0  \\   \\
\realp{\left( \Phi \right)} = 0
\ .
\end{array}
\right.
\label{cond}
\ee
The second condition is again in contrast with the solitonic state
(\ref{kink}), but (\ref{condzero}) is compatible with Lorentz
invariance in five and four dimensions.
\par
There is also a third possible definition which involves a change
of sign of the coordinate $y$ \cite{thirring}:
\be
C_{5(y)}\,\left[\Psi\right](\bol{x},y)
=\left[C\otimes\id_2\right]\,\Psi^\star(\bol{x},-y)
\ .
\ee
Invariance of Eq.~(\ref{dirac+doubling}) now requires
\be
\Phi(\bol{x},y)=\Phi^{\star}(\bol{x},-y)
\ ,
\label{cond2}
\ee
and is not compatible with the $Z_2$ symmetry if we take $\Phi$
real.
\par
To summarize, there appear to be no known definition of charge
conjugation in five dimensions which is compatible with
localization given by the use of the scalar field in
Eq.~(\ref{kink}).
\subsubsection{Parity}
We define the five-dimensional parity (P$_5$) as the usual parity
(P) acting on each family $\psi_i$,
\be
P_5\,\left[\Psi\right](t,{\vec x},y)=
\left[P \otimes\id_2\right]\,\Psi(t,-{\vec x},-y)
\ ,
\label{P5}
\ee
where $P=i\,\gamma^0$.
Invariance of Eq.~(\ref{dirac+doubling}) requires
\be
\Phi(t,\vec x, y)=
\Phi(t,-\vec x, -y)
\ .
\ee
This is in contrast with Eq.~(\ref{kink}) so that the
five-dimensional parity is broken if we choose $\Phi$
kink-like, and the conclusion is the same as for charge conjugation.
\subsubsection{Time reversal}
We define the five-dimensional time reversal (T$_5$) as the usual
time reversal (T) acting on each family $\psi_i$,
\be
T_5\,\left[\Psi\right](t,{\vec x},y)=
\left[T\otimes\id_2\right]\,\Psi^\star(-t,{\vec x},y)
\ ,
\label{T}
\ee
where $T=i\,\gamma^1\,\gamma^3$.
Invariance of Eq.~(\ref{dirac+doubling}) requires
\be
\Phi(t,\vec x,y)=\Phi^\star(-t,\vec x,y)
\ .
\ee
If $\Phi$ is given by Eq.~(\ref{kink}), this symmetry is satisfied.
\subsubsection{CPT}
One expects that, if the five-dimensional theory is a genuine (locally
Lorentz invariant) field theory, CPT holds \cite{jost}.
However, we have three ways of defining CPT depending on the choice
of the operator of charge conjugation and we would like to pick up
the one which is compatible with the localization mechanism given
by the soliton in Eq.~(\ref{kink}) and Eq.~(\ref{localization}).
\par
The preferred CPT is thus given by the combination C$_5$P$_5$T$_5$.
In fact, invariance of Eq.~(\ref{dirac+doubling}) under the
transformation
\be
\left(C_{5}\,P_5\,T_5 \right)[\Psi](x^A)
=\left[\gamma^5\otimes\id_2\right]
\Psi^\star(-x^A)
\ ,
\ee
requires:
\be
&&\left\{ i\,\left[\gamma^4\otimes\id_2\right]\,
\left(D_y+\omega\right)
\phantom{\vec a\over\vec b}
\right.
\nonumber \\
&&\left.
\,
-\left[\id_4\otimes\sigma^3\right]\,
{\Phi(x^A)-\Phi(-x^A)\over 2}\right\}\,\Psi=0
\ .
\label{CPT_5}
\ee
If $\Phi $ is kink-like the latter equation becomes precisely
Eq.~(\ref{localization}) for the localization.
Thus we conclude that localization on the brane is a consequence
of CPT invariance of the (effective) five-dimensional field
theory.
However, since the condition (\ref{CPT_5}) suffers of the same
problem as Eq.~(\ref{cC5}), it violates (local) five-dimensional
Lorentz invariance (while preserving four-dimensional Lorentz
invariance).
This is not surprising, since confined states are not invariant
under rotations and boosts which involve the fifth dimension or,
put another way, the brane itself represents a preferred frame.
Note that separately C$_5$ and P$_5$ are broken but together
with T$_5$ they give Eq.~(\ref{CPT_5}) that becomes
(\ref{localization}) when (\ref{kink}) [and (\ref{vanishing})
for zero modes] are imposed.
\par
For completeness we also consider the remaining cases.
In particular, invariance of Eq.~(\ref{dirac+doubling}) under
the combination $\tilde{\rm C}_5$P$_5$T$_5$,
\be
\left(\tilde C_{5}\,P_5\,T_5\right)[\Psi](x^A)
=-\left[\id_4 \otimes \id_2 \right]
\Psi^{\star}(-x^A)
\ ,
\ee
 requires
\be
&&\left\{\left[\id_4\otimes\sigma^3\right]\,
{\Phi(x^A)+\Phi(-x^A)\over 2}\right.
\nonumber \\
&&\left.\phantom{\vec a\over\vec b}
+{\mu\over a}\,\left[\id_4 \otimes \sigma^1 \right]\right\}\,\Psi=0
\ .
\label{tildeCPT_5}
\ee
If we take $\Phi$ to be given as in Eq.~(\ref{kink}) we have
to set $\mu=0$.
This confirms that $\tilde{\rm C}_5$ is a bad definition of charge
conjugation if we want massive fermions on the brane, although this
choice is fully compatible with five-dimensional Lorentz symmetry.
\par
Finally, invariance of Eq.~(\ref{dirac+doubling}) under the
combination which includes C$_{5(y)}$,
\be
\left(C_{5(y)}\,P_5\,T_5\right)[\Psi](\bol x , y)
=\left[\gamma^5\otimes \id_2 \right]
\Psi ^{\star}(-\bol x , y)
\ ,
\label{defC_yPT_5}
\ee
requires
\be
\Phi(\bol{x},y)=\Phi(-\bol{x},y)
\ .
\label{C_yPT_5}
\ee
This constraint is satisfied by Eq.~(\ref{kink}).
We will see below that this definition of CPT is really four-dimensional.
\subsubsection{CP and T}
Since T$_5$ holds for the choice in Eq.~(\ref{kink}), one expects
that CP is unbroken as a consequence of CPT invariance.
We have three different ways of defining CP depending on the
three possible choices of charge conjugation.
Invariance of Eq.~(\ref{dirac+doubling}) under C$_5$ and P$_5$,
\be
\left(C_5\,P_5\right)[\Psi](x^A)
=-\left[\gamma^2\,\gamma^0 \otimes \id_2 \right]
\Psi ^{\star}(t, -\vec x, -y)
\ ,
\ee
requires:
\be
&&\left\{i\,\left[\gamma^4\otimes\id_2\right]\,
\left(D_y+\omega\right)\phantom{\vec a\over\vec b}\right.
\nonumber \\
&&\left.
\
-\left[\id_4\otimes\sigma^3\right]\,
{\Phi(t,\vec{x},y)-\Phi^{\star}(t,-\vec{x},-y)\over 2}\right\}
\,\Psi=0
\ .
\label{CP_5}
\ee
If $\Phi$ is given by Eq.~(\ref{kink}) the above condition again
becomes the equation for localization and CP is unbroken for
fermions living on the brane (while five-dimensional Lorentz
invariance is violated).
\par
The second possible definition makes use of $\tilde{\rm C}_5$ and
P$_5$:
\be
\left(\tilde C_5\,P_5\right)[\Psi](x^A)
=-\left[\gamma^2\,\gamma^5\,\gamma^0 \otimes \id_2 \right]
\Psi^{\star}(t,-\vec x,-y)
\ .
\ee
Invariance of Eq.~(\ref{dirac+doubling}) then requires:
\be
&&\left\{ \,\left[\id_4\otimes\sigma^3\right]\,{1\over 2}\,
\left(\Phi(t,\vec{x},y)+\Phi^{\star}(t,-\vec{x},-y)\right) \right.
\nonumber \\
&& \left.\phantom{\vec a\over\vec b}
+{\mu \over a}\,\left[ \id_4 \otimes \sigma^1\right]\right\}\,\Psi=0
\ .
\label{tildeCP_5}
\ee
Note that if $\Phi$ is kink-like the first term vanishes and the
above condition is verified for every $\Psi$ if $\mu = 0$.
We conclude that $\tilde{\rm C}_5$P$_5$ respects five-dimensional
Lorentz invariance and is satisfied if there is
no interaction term between the two families of fermions.
However, as we mentioned, this term is necessary to obtain the
four-dimensional mass term and to recover the low energy
phenomenology.
\par
We can also define CP using C$_{5(y)}$ and P$_5$:
\be
\left(C_{5(y)}\,P_5\right)[\Psi](x^A)
=-\left[\gamma^2\,\gamma^0 \otimes \id_2 \right]
\Psi^{\star}(t, -\vec x, y)
\ .
\ee
In this case invariance of Eq.~(\ref{dirac+doubling}) yields:
\be
\Phi (t,\vec x,y) = \Phi^{\star}(t,-\vec x,y)
\ .
\label{CP_y}
\ee
This condition is satisfied if $\Phi$ is given by Eq.~(\ref{kink})
and, as we show below, coincides with the condition that ensures CP
in four dimensions.
\subsection{Four-dimensional symmetries}
\label{four}
In this Section we define discrete symmetries as acting on
eight-spinors as seen from the four-dimensional point of view.
In other words we are looking for operators that act on
$\psi_{1,L}$ and $\psi_{2,R}$ ($\psi_{2,L}$ and $\psi_{1,R}$)
as if they belonged to the same Dirac spinor $\psi^{(1)}$
($\psi^{(2)}$) and are not substantially affected
by the existence of a fifth coordinate, hence they must be
represented by matrices of the form
\be
\left[\begin{array}{ccc}
F^{(1)}_{2\times 2} & 0_{2\times 4} & F^{(1)}_{2\times 2}
\\
0_{4\times 2} & F^{(2)}_{4\times 4} & 0_{4\times 2}
\\
F^{(1)}_{2\times 2} & 0_{2\times 4} & F^{(1)}_{2\times 2}
\end{array}\right]
\ ,
\label{4Dm}
\ee
in which upper indeces denote blocks acting on the corresponding
spinor $\psi^{(i)}$.
\par
Strictly speaking such symmetries are not as fundamental as the
five-dimensional analogues and are properly defined only for
states localized on the brane.
Further, they all naively respect four-dimensional Lorentz
invariance, since four-dimensional Lorentz transformations
must be represented by matrices of the form (\ref{4Dm}) and,
as we show below, C, P and T coincide with the standard
four-dimensional expressions.
We finally note that four-dimensional Lorentz invariance for
confined states also follows from the effective action
(\ref{reduced}).
\subsubsection{Charge conjugation}
We define the four-dimensional charge conjugation as
\be
C_4\,\left[\Psi\right](x^A)
=\left[C\otimes\sigma^1\right]\,\Psi^\star(x^A)
\ee
where, as before, $C=i\,\gamma^2$.
Invariance of Eq.~(\ref{dirac+doubling}) now requires
\be
\left\{i\,\left[\gamma^4\otimes\id_2\right]
\left(D_y+\omega\right)
-\left[\id_4\otimes\sigma^3\right]\,\realp(\Phi)
\right\}\Psi=0
\ .
\ee
If we choose (\ref{kink}), this symmetry is satisfied if the
equation for localization (\ref{localization}) holds.
\par
As in five dimensions, this is not the
only way to define this operation.
In fact we can also define the four-dimensional charge conjugation
as
\be
\tilde C_4\,\left[\Psi\right](x^A)
=\left[\tilde C\otimes \sigma^1\right]\,\Psi^\star(x^A)
\ee
where, as before, $\tilde C=i\,\gamma^2\,\gamma^5$.
Invariance of Eq.~(\ref{dirac+doubling}) requires Eq.~(\ref{condzero})
and we obtain the same results as in five dimensions
[see Eq.~(\ref{cond})].
\par
Analogously to the five-dimensional case, we could also
define charge conjugation as
\be
C_{4(y)}\,\left[\Psi\right](\bol{x},y)=
\left[C\otimes\sigma^1\right]\,\Psi^\star(\bol{x},-y)
\ .
\ee
Invariance of Eq.~(\ref{dirac+doubling}) then requires (\ref{cond2}).
\par
To conclude, while C$_4$-invariance is compatible with localization, both
$\tilde {\rm C}_4$ and C$_{4(y)}$ lead to the same conditions as
the anologous $\tilde {\rm C}_5$ and C$_{5(y)}$ and are therefore
in contrast with localization.
If one wants an operator of charge conjugation which, from
a four-dimensional point of view, gives rise to an invariance of
Eq.~(\ref{dirac+doubling}), one is then forced to discard
$\tilde {\rm C}_4$ and C$_{4(y)}$ as candidates for charge
conjugation.
\subsubsection{Parity}
We define the four-dimensional parity as follows
\be
P_4\,\left[\Psi\right](t,{\vec x},y)
=\left[P\otimes\sigma^1\right]\,
\Psi(t,-{\vec x},y)
\ ,
\ee
where, as before, $P=i\,\gamma^0$.
Invariance of Eq.~(\ref{dirac+doubling}) requires
\be
&&\left\{i\left[\gamma^4\otimes\id_2\right]
\left(D_y+\omega\right)
\phantom{1\over 2}\right.
\nonumber \\
&&\left.\phantom{\left\{\right.}
-\left[\id_4\otimes\sigma^3\right]
\,{\Phi(t,\vec x,y)+\Phi(t,-\vec x,y)\over 2}
\right\}\Psi=0
\ .
\ee
One more time, if $\Phi$ is given by (\ref{kink}), we get that $P$
is satisfied when the equation for localization holds.
\subsubsection{Time reversal}
The four-dimensional definition of this operation T$_4$ is exactly
the same as T$_5$ in five dimensions.
\subsubsection{CPT}
Invariance of Eq.~(\ref{dirac+doubling}) under the combination
\begin{eqnarray}
\left(C_{4}\,P_4\,T_4 \right) [\Psi](\bol{x},y)
&=&\left[\gamma^5 \otimes \id_2 \right]
\Psi^{\star}(-\bol x , y) \nonumber \\
&=&\left(C_{5(y)}\,P_5\,T_5\right)[\Psi](\bol{x},y)
\label{defCPT_4}
\end{eqnarray}
requires Eq.~(\ref{C_yPT_5}).
If $\Phi$ does not depend on $\bol{x}$ (as the kink) the latter
equation is automatically satisfied.
\subsubsection{CP and T}
Invariance of Eq.~(\ref{dirac+doubling}) under combination of the
previous operations C$_4$ and P$_4$,
\begin{eqnarray}
\left(C_{4}\,P_4\right)[\Psi] (x^A)
&=&-\left[\gamma^2\,\gamma^0\otimes \id_2 \right]
\Psi^{\star}(t,- \vec x , y) \nonumber \\
&=&\left(C_{5(y)}\,P_5\right) [\Psi] (x^A)
\ ,
\label{defCP_4}
\end{eqnarray}
requires Eq.~(\ref{CP_y}).
\section{Conclusions}
\label{conclusions}
We have analyzed discrete symmetries in a four-dimensional
brane-world of codimension one and their relation with
localization and Lorentz invariance.
We found that the usual way of localizing fermions with
a (anti-) kink \cite{rubakov} is compatible with CPT
in five dimensions (and follows from it), provided one
uses C$_5$, P$_5$ and T$_5$ given respectively in
Eqs.~(\ref{C5}), (\ref{P5}) and (\ref{T}).
In fact, we have shown that, among the three expressions of
CPT that we have considered in five dimensions, just
C$_5$P$_5$T$_5$ is consistent with a brane-world.
On the contrary, $\tilde {\rm C}_5$P$_5$T$_5$
is not compatible with a mass term [such as the vacuum
expectation value of the Higgs field, $\mu$ in
Eq.~(\ref{5Daction+doubling})] for fermions on
the brane and C$_{5(y)}$P$_5$T$_5$ is actually
four-dimensional (since it equals C$_4$P$_4$T$_4$).
Correspondingly, CPT in four dimensions ({\em i.e.},
C$_4$P$_4$T$_4$, which is properly defined only for localized
fermions) does not impose any further restriction except
Eq.~(\ref{C_yPT_5}).
It is also possible to obtain Eq.~(\ref{localization}) for
the localization from a purely four-dimensional point of view:
it is sufficient to impose the invariance of the Dirac
equation (\ref{dirac+doubling}) under either C$_4$ or P$_4$.
\par
It is nevertheless worth noting that it is
$\tilde {\rm C}_5$P$_5$T$_5$ which respects five-dimensional
(local) Lorentz invariance.
Instead, C$_5$P$_5$T$_5$ breaks this symmetry (while preserving
Lorentz invariance on the brane), as one could naively
expect since the brane represents a preferred frame.
Since Lorentz invariance is a main hypothesis for the CPT
theorem (in any dimensions) \cite{jost}, this seems to play
in favour of the expression involving $\tilde{\rm C}_5$
and one would therefore expect that five-dimensional CPT is
broken when there are massive fermions on the brane.
Whether the breakdown of this five-dimensional symmetry leads
to observable effects in our (brane-)world will be the subject
of subsequent investigations.
\par
Since the origin of the reference frame on the brane
is arbitrary, the condition (\ref{C_yPT_5}) for CPT invariance
in four dimensions implies that the kink must be static
and homogeneous, although such a condition is not required
for CPT invariance from the five-dimensional point of view.
One would expect that background fields, such as $\Phi$, are
not static in a cosmological context (at least during the
early stages of the Universe) and can therefore be a source
of CPT asymmetries on the brane-world.
Analogously, spatial inhomogeneities could generate local
violations of CP in four dimensions.
It could be that some work can be done in that direction.
\acknowledgments
We would like to thank G.~Venturi for useful discussions.
\appendix
\section{Thick brane}
\label{thick}
In this Section we explicitly consider the case of a thick brane.
For this purpose it is useful to approximate the kink as
\cite{schmaltz,g2}
\be
\Phi=\left\{\begin{array}{ll}
-(m_f^2/2)\,L_f &\ \ \ \ \ \ \ y<-L_f
\\
(m_f^2/2)\,y &\ \ \ \ \ \ \ |y|<L_f
\\
+(m_f^2/2)\,L_f &\ \ \ \ \ \ \ y>+L_f
\ ,
\end{array}\right.
\ee
where $L_f$ is the typical brane thickness for fermions, and take
$a(y)\sim 1$ for $|y|<L_f$.
It is then easy to show that fermions have a confined massless
(chiral) mode, together with a tower of states (which are allowed
only if their mass is smaller than the threshold
$m_f^2\,L_f/2$ \cite{g2}).
Since $\gamma^5=\Pi_L-\Pi_R$ (the difference between L and R
chiral projectors), one can introduce ``creation and annihilation''
operators
\be
\ad=-{1\over m_f}\,\left(\partial_y-\Phi\right)
\ , \ \ \ \ \
\a={1\over m_f}\,\left(\partial_y+\Phi\right)
\ ,
\ee
such that $[\a,\ad]=1$ and the Lagrangian density for $|y|<L_f$
becomes
\be
L_{(5)}=
\bar\Psi\,\left(i\,\Ds-m_f\,\a\,\Pi_L-m_f\,\ad\,\Pi_R\right)\,\Psi
\ .
\ee
This allows an expansion for the fermions
\be
&&\Psi(\bol{x},y)=H_0(y)\,\Pi_L\,\psi^{(0)}(\bol{x})
\nonumber \\
&&+\sum_{n=1}^{N_f}
\left[H_{n}(y)\,\Pi_L+H_{n-1}(y)\,\Pi_R\right]\psi^{(n)}(\bol{x})
\ ,
\label{f_exp}
\ee
where $H_{n}$ are the normalized eigenfunctions of the harmonic
oscillator.
From the above it is clear that while the zero mode is a two-component
spinor, modes with $n\ge 1$ can have both chiralities.
\par
Since the zero mode is massless (because $\a\,H_0=0$),
\be
(\ps+m_f\,\a)\,H_0\,\Pi_L\,\psi^{(0)}
=H_0\,\ps\,\Pi_L\,\psi^{(0)}=0
\ ,
\ee
$\psi^{(0)}$ can be taken as a L Weyl spinor,
$\psi^{(0)}=\Pi_L\,\psi^{(0)}$.
We note in passing that $\a\,H_0=0$ is precisely the equation
which ensures the confinement of the left zero mode within
a width $\ell_f\sim 1/m_f$ around $y=0$.
Since for the (would-be) right zero mode the corresponding
equation $\ad\,\bar R_0=0$ does not admit any (non-vanishing)
normalizable solution in $y\in\real$, we have set
$\Pi_R\,\psi^{(0)}=0$.
\par
\begin{figure}
\centerline{\epsfxsize=3.0in
\epsfbox{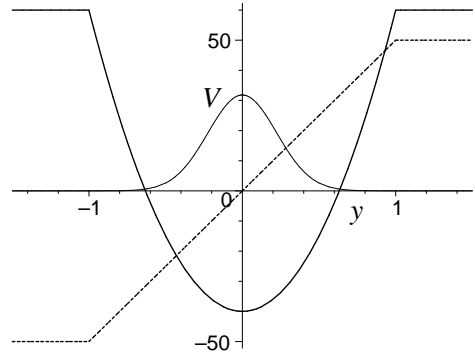}}
\caption{Sketch of the scalar field $\Phi$ (magnified by a factor
of 5; dashed line) and the corresponding confining potential in
Eq.~(\ref{KG}) (continuous line) for $m_f^2/2=10$ (in units with
$L=1$).
The Gaussian curve represents the ground state $H_0$ (magnified by
a factor of 10).}
\label{g2plot01}
\end{figure}

The sum in Eq.~(\ref{f_exp}) ends with a maximum integer
$N_f<\infty$.
The reason for such a cut-off can be easily understood if we
set $\As=0$ and write down the Klein-Gordon equation corresponding
to the Dirac equation obtained from $S$,
\be
&&\left[\ps-m_f\,\left(\ad\,\Pi_L-\a\,\Pi_R\right)\right]
\left[\ps+m_f\,\left(\a\,\Pi_L+\ad\,\Pi_R\right)\right]\Psi
\nonumber \\
&&=-\left[p^2+m_f^2\,\left(\ad\,\a\,\Pi_L+\a\,\ad\,\Pi_R\right)
\right]
\Psi
\nonumber \\
&&=
-\left[p^2+\left(-\partial_y^2+\Phi^2\right)\right]\,
\Psi=0
\ .
\label{KG}
\ee
It is thus clear that only those modes $\psi_n$ whose eigenvalues
$m_f^2\,n<\Phi^2(L)\equiv M_f^2$ can be retained
(see Fig.~\ref{g2plot01}).
\par

\begin{references}
%
%
\bibitem[a]{email}E-mail: casadio@bo.infn.it
\bibitem[b]{email}E-mail: gruppuso@bo.infn.it
%
%
\bibitem{kaluza}
Th. Kaluza, Sitz. Preuss. Akad., 966 (1921).
%
\bibitem{thirring}
W. Thirring, Acta Phys. Austr. Suppl. IX, 256 (1972).
%
\bibitem{green}
M.B. Green, J.H. Schwarz and E. Witten, {\em Superstring theory}
(Cambridge Univ. Press, Cambridge, England, 1987).
%
\bibitem{horava}
P. Horava and E. Witten, Nucl. Phys. {\bf B460}, 506 (1996);
Nucl. Phys. {\bf B475}, 94 (1996); E. Witten, Nucl. Phys. {\bf B471},
135 (1996).
%
\bibitem{polchinski}
J. Polchinski, {\it TASI Lectures on D-branes}, hep-th\-/\-97\-02\-136.
%
\bibitem{arkani}
N. Arkani-Hamed, S. Dimopoulos and G. Dvali, Phys. Lett. B {\bf 429},
263 (1998); Phys. Rev. D {\bf 59}, 0860004 (1999).
%
\bibitem{randall}
L. Randall and R. Sundrum, Phys. Rev. Lett. {\bf 83}, 3370 (1999);
{\bf 83}, 4690 (1999).
%
\bibitem{long}
J.C. Long, H.W. Chan and J.C. Price, Nucl. Phys. {\bf B539}, 23 (1999).
%
\bibitem{giudice}
G.F. Giudice, R. Rattazzi and J.W. Wells, Nucl. Phys. {\bf B544}, 3
(1999).
%
\bibitem{rubakov}
V.A. Rubakov and M. Shaposnikov, Phys. Lett. {\bf B 125}, 136 (1983);
M. Visser,  Phys. Lett. {\bf B 159}, 22 (1985);
M. Gogberashvili, Europhys. Lett. {\bf 49}, 396 (2000).
%
\bibitem{schmaltz}
N. Arkani-Hamed and M. Schmaltz, Phys. Rev. D {\bf 61} 033005 (2000).
%
\bibitem{lykken}
J. Lykken, R.C. Myers and J. Wang, hep-th/0006191.
%
\bibitem{g2}
R. Casadio, A. Gruppuso and G. Venturi, Phys. Lett. {\bf B 495}
(2000) 378.
%
\bibitem{graesser}
M.L. Graesser, Phys. Rev. D {\bf 61}, 074019 (2000).
%
\bibitem{kink}
We note that the (anti-)kink is not periodic in $y$ and therefore,
although it matches the Z$_2$ symmetry, it would break the
(orbifold) $S^1/$Z$_2$ symmetry.
However, since it is the shape of $\Phi$ for $|y|$ less then
(a few times) $1/p$ which induces the confinement, one can make
it periodic by modifying $\Phi$ at larger values of $|y|$
without substantially affecting this property.
Alternatively, one can just take the limit which yields the
periodic form in Eq.~(\ref{kink2}).
%
\bibitem{rubakov2}
S.L. Dubovsky, V.A. Rubakov and P.G. Tinyakov, Phys. Rev. D {\bf 62} 105011
(2000).
%
\bibitem{gauge}
It is always possible to set $A_4=0$ as a choice of gauge, however
we shall not make any restriction.
%
\bibitem{peskin}
M.E. Peskin and D.V. Schroeder, {\em An introduction to quantum
field theory} (Perseus Books, Cambridge, Massachusetts, 1995).
%
\bibitem{pais}
A. Pais, J. Math. Phys. {\bf 6}, 1135 (1962).
%
\bibitem{jost}
R. Jost, {\em The general theory of quantized fields} (American
Mathematical Society, Providence, Rhode Island, 1965).
%
\end{references}
\end{document}